\documentclass[floats,floatfix,showpacs,preprintnumbers,amssymb,prl,twocolumn,superscriptaddress,nolongbibliography,reprint]{revtex4-2}

\usepackage{amssymb,amsmath,verbatim,mathtools,needspace,enumitem,etoolbox,graphicx,physics,microtype,afterpage,xspace,tabularx,lmodern,multirow}
\usepackage{gensymb}
\usepackage[normalem]{ulem}
\usepackage[dvipsnames, usenames]{xcolor}
\usepackage{xr-hyper}
\definecolor{linkcolor}{rgb}{0.0,0.3,0.5}
\usepackage[unicode, colorlinks=true, linkcolor=linkcolor, citecolor=linkcolor, filecolor=linkcolor, urlcolor=linkcolor, linktocpage, breaklinks]{hyperref}
\usepackage[all]{hypcap}
\usepackage[T1]{fontenc}
\usepackage[utf8]{inputenc}
\usepackage[export]{adjustbox}
\usepackage[usenames,dvipsnames]{xcolor}
\hypersetup{colorlinks=true,citecolor=romared,linkcolor=romared,urlcolor=romared}

\definecolor{romared}{RGB}{142,0,28}

\newcommand{\be}{\begin{equation}}
\newcommand{\ee}{\end{equation}}

\def\be{\begin{equation}}
\def\ee{\end{equation}}
\newcommand{\beq}{\begin{eqnarray}}
\newcommand{\eeq}{\end{eqnarray}}

\usepackage{makecell}
\usepackage{soul}

\newcolumntype{Y}{>{\centering\arraybackslash}X}

\usepackage{soul}

\definecolor{romared}{RGB}{142,0,28}

\makeatletter
\newcommand*{\addFileDependency}[1]{
  \typeout{(#1)}
  \@addtofilelist{#1}
  \IfFileExists{#1}{}{\typeout{No file #1.}}
}
\makeatother

\begin{document}
\title{Dark matter and modified gravity: Einstein clusters from a non-minimally coupled vector field}
\author{Pedro G. S. Fernandes}
\affiliation{Institut f\"ur Theoretische Physik, Universit\"at Heidelberg, Philosophenweg 12, 69120 Heidelberg, Germany}
\author{Vitor Cardoso} 
\affiliation{Center of Gravity, Niels Bohr Institute, Blegdamsvej 17, 2100 Copenhagen, Denmark}
\affiliation{CENTRA, Departamento de F\'{\i}sica, Instituto Superior T\'ecnico -- IST, Universidade de Lisboa -- UL, Avenida Rovisco Pais 1, 1049-001 Lisboa, Portugal}

\date{\today}

\begin{abstract}
We show that a vector field non-minimally coupled to gravity reproduces \emph{exactly} the dynamics of an Einstein cluster -- a large ensemble of non-interacting particles moving on circular geodesics under their collective gravitational field. Since Einstein clusters are known to be able to account for flat galactic rotation curves, our results suggest that such rotation curves may arise as a manifestation of modified gravity.
\end{abstract}

\maketitle

\noindent \textbf{\textit{Introduction.}}
Although General Relativity (GR) has been extremely successful in describing gravitational phenomena across a wide range of scales, it cannot describe all observed gravitational phenomena using visible matter alone. In particular, the dynamics of galaxies, galaxy clusters, and large-scale structure reveal systematic discrepancies between the gravitational field inferred from luminous baryonic matter and that required by observation~\cite{Zwicky:1933gu,Einasto:1974dra,Rubin:1978kmz,Rubin:1980zd,Tyson:1990yt,Wittman:2000tc}. Measurements of galactic rotation curves show that orbital velocities remain approximately flat at large radii rather than decreasing according to the Newtonian expectation, while cluster dynamics and gravitational lensing indicate substantially more gravitating mass than can be accounted for by stars, gas, and dust. These anomalies are conventionally attributed to an additional non-luminous component known as \emph{dark matter}, which in the standard cosmological framework is modeled as a cold, pressureless fluid interacting predominantly through gravity. Despite its success in explaining a broad range of astrophysical and cosmological observations, the fundamental nature of dark matter remains unknown~\cite{Freese:2008cz,Navarro:1995iw,Clowe:2006eq,Bertone:2004pz,Kahlhoefer:2017dnp,PerezdelosHeros:2020qyt}, even though a vast array of dark matter candidates has been proposed, ranging from new fundamental fields to primordial black holes \cite{Cirelli:2024ssz,Arbey:2021gdg,Oks:2021hef,Carr:2016drx,Green:2024bam,Carr:2020xqk}.

An alternative to the standard dark matter paradigm is that the observed discrepancies arise not from an unseen matter component, but from a modification of gravitational dynamics on galactic or cosmological scales. In this view, the laws governing gravity are modified in the regimes where the dark matter hypothesis is usually invoked, and many such proposals are broadly associated with modified Newtonian dynamics (MOND)~\cite{1983ApJ...270..365M,1983ApJ...270..371M,1983ApJ...270..384M} and its relativistic extensions, see e.g.~\cite{1984ApJ...286....7B,Sanders:1996wk,PhysRevD.70.083509,Zlosnik:2006zu,Babichev:2011kq,Khoury:2014tka,Skordis:2020eui}. The distinction between introducing new matter and modifying gravity, however, is subtle. By Lovelock's theorem~\cite{Lovelock:1971yv}, any consistent extension of gravitational dynamics beyond standard GR requires additional fields or degrees of freedom beyond the metric itself. These extra gravitational degrees of freedom contribute to the effective stress-energy content of spacetime and can therefore behave phenomenologically as an additional dark sector. In this sense, many modified gravity theories may equally be interpreted as theories in which new gravitational fields mimic dark matter rather than eliminate the need for it.

In this letter, we introduce a vector–tensor theory in which the additional gravitational degrees of freedom can behave \emph{exactly} like an Einstein cluster~\cite{Einstein:1938yz} -- a large ensemble of non-interacting particles moving on circular geodesics under their collective gravitational field. Therefore, the modified gravity sector is able to generate the same gravitational effects ordinarily attributed to dark matter in galactic systems. In this way, the vector field does not merely modify the gravitational field equations, but dynamically realizes an effective dark component whose energy density profile can explain galactic-scale observations.

We work in units $c=G=1$.

\noindent \textbf{\textit{The Einstein cluster.}}
The Einstein cluster~\cite{Einstein:1938yz} is a model that describes a large collection of non-interacting particles moving along circular geodesics under their collective gravitational field. The resulting system is spherically symmetric and centrifugally stable, with the outward centrifugal force on each particle balanced by the inward pull of gravity. Einstein clusters have been extensively studied~\cite{hoganReconstructionMinkowskianSpacetime1978,1968ApJ...153L.163Z,1974RSPSA.337..529F,Comer:1993rx,kumardattaNonstaticSphericallySymmetric1970,bondiDattasSphericallySymmetric1971,Gair:2001qu,Szybka:2018hoe,Mahajan:2007vw,Magli:1997qf}, and proposed as models for dark matter, as they can reproduce galactic rotation curves~\cite{Acharyya:2023rnq,Jusufi:2022jxu,Boehmer:2007az,Lake:2006pp,Geralico:2012jt}. Also, Einstein clusters have recently garnered significant attention as a framework for modelling black holes embedded within dark matter halos~\cite{Cardoso:2021wlq,Figueiredo:2023gas,Speeney:2024mas,Cardoso:2022whc,Jusufi:2022jxu,Shen:2023erj,Pezzella:2024tkf,Maeda:2024tsg,Spieksma:2024voy,Konoplya:2022hbl,Macedo:2024qky,Xavier:2023exm,Dai:2023cft,Konoplya:2021ube,Ovgun:2025bol,Myung:2024tkz,Gliorio:2025cbh,Fernandes:2025osu}.

The Einstein cluster is described by the generic static and spherically symmetric line-element
\begin{equation}
    \dd s^2 = -e^{2\phi(r)} \dd t^2 + \frac{\dd r^2}{1-\frac{2m(r)}{r}} + r^2 \left( \dd \theta^2 + \sin^2 \theta \dd \varphi^2 \right),
    \label{eq:metric}
\end{equation}
together with the effective stress-energy tensor (see e.g. Ref. \cite{Acharyya:2023rnq} for a detailed derivation)
\begin{equation}
    T^{\mu}_{\phantom{\mu} \nu} = \mathrm{diag}\left(-\rho(r), 0, p_t(r),p_t(r)\right),
    \label{eq:stress-energy}
\end{equation}
where $\rho$ and $p_t$ are the energy density and the tangential pressure of the cluster, respectively. Note that the radial pressure, $T^{r}_{\phantom{r} r}$, is vanishing. The Einstein equations and the Bianchi identities for this system impose
\begin{equation}
    m = \frac{r^2 \phi'}{1+2r \phi'},
    \label{eq:relationmetric}
\end{equation}
\begin{equation}
    m' = 4\pi r^2 \rho,
    \label{eq:EinCluster2}
\end{equation}
\begin{equation}
    p_t = \frac{1}{2} r \phi' \rho,
    \label{eq:EinCluster3}
\end{equation}
where the prime denotes a derivative with respect to $r$. The system can be solved once a profile for $\phi$, $m$, $\rho$ or $p_t$ is chosen. The metric potential $\phi$ can be inferred from observations of galactic rotation curves, see Ref.~\cite{Lake:2003tr}. Due to this freedom in choosing the correct profile to match observations, the Einstein cluster is able to reproduce observed galactic rotation curves \cite{Acharyya:2023rnq,Jusufi:2022jxu,Boehmer:2007az,Lake:2006pp,Geralico:2012jt}, assuming dark matter has an effective stress-energy tensor given by Eq.~\eqref{eq:stress-energy}. However, this model offers little insight into the fundamental nature of dark matter.

\noindent \textbf{\textit{The theory.}} 
The main goal of this work is to show that a vector field non-minimally coupled to gravity can exactly reproduce the dynamics of an Einstein cluster, \ul{retaining the freedom in choosing the metric potential $\phi$ to match observational data as a built-in feature}.
We consider a theory of gravity coupled to a vector field $V_\mu$ given by
\begin{equation}
	S = \int \dd^4 x \sqrt{-g}\left( \frac{1}{16\pi} R - \frac{1}{4} F_{\mu \nu}F^{\mu \nu} + \gamma\, G_{\mu \nu} V^\mu V^\nu \right),
    \label{eq:theory}
\end{equation}
where $F_{\mu \nu} = \partial_\mu V_\nu - \partial_\nu V_\mu$ is the kinetic term of the vector field, $G_{\mu \nu}$ is the Einstein tensor, $\gamma$ is a dimensionless coupling constant. From an effective field theory perspective, this theory can be motivated as the most general vector-tensor theory, up to mass dimension four, where: parity symmetry is imposed on the vector field ($V_\mu \to -V_\mu$); $U(1)$ gauge invariance is broken solely through couplings to gravity; the equations of motion remain second order and the theory belongs to the generalized Proca class \cite{Heisenberg:2014rta}, ensuring no pathological degrees of freedom propagate.
Higher-order operators would be suppressed by some UV energy scale, and can safely be neglected at low energies.
The vector field is not identified with the photon, but rather with a new field yet to be directly detected.
Some black hole solutions and other astrophysical objects of the theory \eqref{eq:theory} have previously been explored in Refs. \cite{Chagoya:2016aar,Babichev:2017rti,Minamitsuji:2016ydr,Tasinato:2022vop,Chagoya:2023ddb,Heisenberg:2017xda,Heisenberg:2017hwb,Tasinato:2022vop,Bakopoulos:2024zke,Geng:2015kvs}.

The field equations of the theory~\eqref{eq:theory} are
\begin{equation}
    G_{\mu \nu} = 8\pi T_{\mu \nu},
    \label{eq:EinEqs}
\end{equation}
and
\begin{equation}
	\nabla_\nu F^{\mu \nu} = 2 \gamma G_{\phantom{\mu}\nu}^{\mu} V^\nu,
    \label{eq:VFEqs}
\end{equation}
where the stress-energy tensor of the vector field is
\begin{widetext}
\begin{equation}
    \begin{aligned}
        T_{\mu \nu} = & F_{\mu}^{\phantom{\mu}\alpha} F_{\nu \alpha} - \frac{1}{4}g_{\mu \nu} F_{\alpha \beta} F^{\alpha \beta} + \gamma \bigg[ V^2 R_{\mu \nu} + V_\mu V_\nu R - 4 V^\alpha R_{\alpha(\mu} V_{\nu)} - \nabla_\mu \nabla_\nu V^2 \\& + 2 \nabla_\alpha \nabla_{(\mu} \left(V_{\nu)}V^\alpha \right) - \Box \left(V_\mu V_\nu \right) + g_{\mu \nu} \left( G_{\alpha \beta} V^\alpha V^\beta + \Box V^2 - \nabla_\alpha \nabla_\beta \left( V^\alpha V^\beta \right)  \right) \bigg],
    \end{aligned}
\end{equation}
\end{widetext}
and where we used the shorthand notation $V^2=V_\alpha V^\alpha$. The vector-field equations~\eqref{eq:VFEqs} are of Maxwell type, but sourced by the spacetime curvature. Consequently, in regions where the curvature is negligible, the vector field $V_\mu$ effectively behaves like a standard U(1) gauge field, recovering the usual gauge-invariant dynamics.

In the regimes where the vector is not excited, $V_\mu=0$, the field equations become those of GR, and therefore all solutions of GR are solutions of this theory with trivial vector. If the vector is not excited, for example, at solar system scales, the theory is automatically compatible with all solar system observations.

\noindent \textbf{\textit{The Einstein cluster from an action principle.}}
We start by considering the static and spherically symmetric line-element given in Eq. \eqref{eq:metric} and the most general ansatz for vector field which is compatible with spherical symmetry
\begin{equation}
    V_\mu \dd x^\mu = v_0(r) \dd t + v_1(r) \dd r.
\end{equation}

The main motivation to consider a coupling $G_{\mu \nu} V^\mu V^\nu$, in this context, is that it leads to Eq.~\eqref{eq:relationmetric}, which is the characteristic relation between metric potentials of the Einstein cluster. This relation follows directly from the vector field equations~\eqref{eq:VFEqs}. In particular, since the only non-vanishing component of $\nabla_\nu F^{\mu \nu}$ is the temporal component (due to gauge-invariance), the radial component of the vector field equations automatically implies $G_{\phantom{r}r}^{r}=0$, if $v_1(r)\neq0$, whose solution is given by Eq.~\eqref{eq:relationmetric}.

Since the radial vector field equation implies $G_{\phantom{r}r}^{r}=0$, we have to solve $T_{\phantom{r}r}^{r}=0$, independently. Fortunately, this equation is algebraic in $v_1(r)$ and can be solved analytically
\begin{equation}
    v_1^2 = e^{-2\phi} r^2 \left( \frac{2v_0^2 \phi'- 2 v_0 v_0'}{r} - \frac{v_0'^2}{4\gamma} \right).
    \label{eq:v1}
\end{equation}
With this profile for the radial component of the vector field, there are only two independent field equations left to solve, which we choose to be the $tt$ component of the Einstein equations, and the temporal component of the vector equation. These equations are linear in the second radial derivatives of $\phi$ and $v_0$, and after combining them in a suitable way, the system can be written as
\begin{equation}
    \begin{aligned}
        &a \left( r e^{2\phi} \right)'' = (1-4\gamma) 8\pi r v_0'^2 \left( r e^{2\phi} \right)',\\&
        a\left( r^2 v_0' \right)' = (1-4\gamma) 4\pi r^2  v_0'^2 \left[ (r v_0)' - (1-4\gamma) v_0 \right].
    \end{aligned}
    \label{eq:system_before_crit_coupling}
\end{equation}
where we defined
\begin{equation}
    a = \left( r e^{2\phi} \right)' - 2\pi \left[ \left( r v_0 \right)'^2 - \left( 1-4\gamma\right) v_0 \left( (1-8\gamma)v_0+2r v_0' \right) \right].
\end{equation}
We were not able to solve the full system analytically for general $\gamma$, due to its complexity and non-linearity~\footnote{There are, however, particular solutions with constant $v_0$ which are of the Schwarzschild type.}.

However, we notice that the system drastically simplifies when
\begin{equation}
    \gamma=\frac{1}{4},
    \label{eq:gamma}
\end{equation}
since the field equations become
\begin{equation}
    \begin{aligned}
        &\left[ \left( r e^{2\phi} \right)' - 2\pi \left(r v_0 \right)'^2 \right] \left( r e^{2\phi} \right)'' = 0,\\&
        \left[ \left( r e^{2\phi} \right)' - 2\pi \left(r v_0 \right)'^2 \right] \left( r^2 v_0' \right)' = 0.
    \end{aligned}
    \label{eq:eqs_crit_coupling}
\end{equation}
From now on, we consider only the theory with $\gamma=1/4$.

For this theory, there are two possible branches of solutions, depending on whether the quantity $a=\left( r e^{2\phi} \right)' - 2\pi \left(r v_0 \right)'^2$ is vanishing or not. 
When it is not vanishing, $a\neq 0$, the only asymptotically flat solution is a Schwarzschild metric together with the vector profile $v_0=Q_1+Q_2/r$, where $Q_1$ and $Q_2$ are integration constants, interpreted as independent charges associated with the vector. Despite the non-trivial profiles, the stress-energy tensor of the vector field is vanishing. This branch of solutions is the \emph{stealth} Schwarzschild solution first found in Ref.~\cite{Chagoya:2016aar}.

The other branch occurs when
\begin{equation}
    \left( r e^{2\phi} \right)' - 2\pi \left(r v_0 \right)'^2 = 0,
    \label{eq:final_relation_derivation}
\end{equation}
which has the formal solution
\begin{equation}
    v_0 = \frac{1}{\sqrt{2\pi} r} \left[ Q \pm \int \sqrt{\left( r e^{2\phi}\right)'} \dd r \right],
\end{equation}
where $Q$ is an integration constant that emerges from the homogeneous solution of Eq.~\eqref{eq:final_relation_derivation}, and is interpreted as a charge associated with the vector field.
In this case all field equations are solved, while we still have the freedom to specify the metric potential $\phi$. In other words, the metric potential $\phi$ is unconstrained by the field equations and fixes all other quantities of the system, $m(r)$, $v_1(r)$, and $v_0(r)$ through the relations given in Eqs.~\eqref{eq:relationmetric}, \eqref{eq:eqs_crit_coupling} (top), and \eqref{eq:final_relation_derivation}, respectively.
This degeneracy is only possible if $\gamma=1/4$, because in any other case, the system~\eqref{eq:system_before_crit_coupling} is composed of two linearly-independent equations that always fully determine $v_0$ and $\phi$.

Since the metric potential $\phi$ remains unconstrained by the field equations, it must instead be fixed by physical considerations and observations. Because it can be inferred from observations of galactic rotation curves \cite{Lake:2003tr}, it can always be fixed, for each independent observation, to reproduce observed galactic rotation curves \cite{Acharyya:2023rnq,Jusufi:2022jxu,Boehmer:2007az,Lake:2006pp,Geralico:2012jt}.
As a result, the effects of dark matter can be identified in this context as the modified gravity effects generated by the vector field, and may vary from galaxy to galaxy, being determined entirely by the dynamical evolution of the Universe.

The stress-energy tensor of an Einstein cluster generally satisfies all standard energy conditions for typical dark matter density profiles. Since the vector-field configuration in our theory exactly reproduces the Einstein cluster stress-energy tensor at the background level, the same conclusions apply: the vector-field stress-energy tensor also respects the energy conditions.

\noindent \textbf{\textit{Black holes in dark matter halos.}}
All recently explored solutions of spherically symmetric black holes surrounded by dark matter halos, modelled as Einstein clusters \cite{Cardoso:2021wlq,Figueiredo:2023gas,Speeney:2024mas,Cardoso:2022whc,Jusufi:2022jxu,Shen:2023erj,Pezzella:2024tkf,Maeda:2024tsg,Spieksma:2024voy,Konoplya:2022hbl,Macedo:2024qky,Xavier:2023exm,Dai:2023cft,Konoplya:2021ube,Ovgun:2025bol,Myung:2024tkz,Gliorio:2025cbh,Fernandes:2025osu}, are solutions of this theory, because the Einstein cluster and the degenerate branch of our theory are fully equivalent.

\begin{figure}[]
	\centering
	\includegraphics[width=\linewidth]{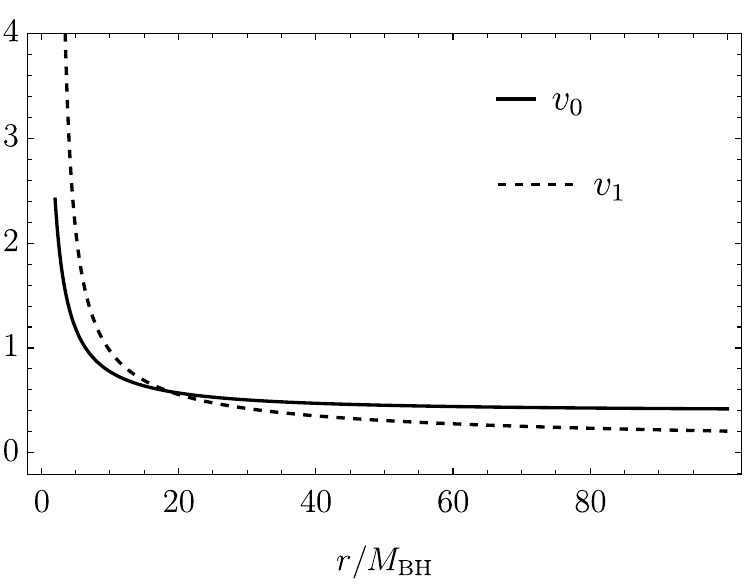}\hfill
    \caption{
    Profiles for the vector field components $v_0$ (solid line) and $v_1$ (dashed line), for the solution presented in Eq.~\eqref{eq:SolHernquist}, with parameters $M_{\rm halo}/M_{\rm BH}=10$, and $a_0/M_{\rm BH}=100$. The initial condition used to solve Eq.~\eqref{eq:final_relation_derivation} is $v_0(2 M_{\rm BH}) \approx 2.42$, because for this value, the homogeneous solution is set to zero, i.e., the profile $v_0$ does not fall-off as $1/r$ but as $1/r^2$.
    }
    \label{fig:vector-profiles}
\end{figure}

To provide an example, we consider here the solution obtained in Ref.~\cite{Cardoso:2021wlq}, which represents a Schwarzschild black hole immersed in a Hernquist dark matter halo, given by
\begin{widetext}
    \begin{equation}
        \begin{aligned}
            &m = M_{\rm BH} + \frac{M_{\rm halo} r^2}{(r+a_0)^2}\left( 1-\frac{2M_{\rm BH}}{r}\right)^2,\\&
            e^{2\phi} = \left( 1-\frac{2M_{\rm BH}}{r}\right) \exp{\sqrt{\frac{M_{\rm halo}}{2a_0-M_{\rm halo}+4 M_{\rm BH}}} \left( 2\arctan \left[ \frac{a_0-M_{\rm halo}+r}{\sqrt{M_{\rm halo}(2a_0-M_{\rm halo}+4 M_{\rm BH})}} \right] - \pi \right) },
        \end{aligned}
        \label{eq:SolHernquist}
    \end{equation}
\end{widetext}
where $M_{\rm BH}$ is the mass of the black hole, $M_{\rm halo}$ the mass of the dark matter halo, and $a_0$ a typical length scale associated with the halo.
These metric functions obey the relation~\eqref{eq:relationmetric}. To demonstrate that this geometry is a solution to the vector-tensor theory~\eqref{eq:theory}, we have numerically obtained the profiles for $v_0$ and $v_1$, by solving Eqs.~\eqref{eq:v1} and \eqref{eq:final_relation_derivation}. The results are presented in Fig.~\ref{fig:vector-profiles}.
We observe that profiles for both components of the vector field that support the geometry~\eqref{eq:SolHernquist} do exist. We have reached similar conclusions for the other geometries of black holes embedded in dark matter halos presented in Refs.~\cite{Cardoso:2021wlq,Figueiredo:2023gas,Speeney:2024mas,Cardoso:2022whc,Jusufi:2022jxu,Shen:2023erj,Pezzella:2024tkf,Maeda:2024tsg,Spieksma:2024voy,Konoplya:2022hbl,Macedo:2024qky,Xavier:2023exm,Dai:2023cft,Konoplya:2021ube,Ovgun:2025bol,Myung:2024tkz,Gliorio:2025cbh}.

\noindent \textbf{\textit{Discussion.}}
We have shown that the theory \eqref{eq:theory} with $\gamma = 1/4$, exactly reproduces the dynamics of an Einstein cluster, with the metric potential $\phi$ remaining freely specifiable. This freedom is physically significant, since $\phi$ can be determined directly from observed galactic rotation curves~\cite{Lake:2003tr}. Consequently, for each individual galactic system, $\phi$ may be chosen so that the resulting solution reproduces the observed rotation profile exactly \cite{Acharyya:2023rnq,Jusufi:2022jxu,Boehmer:2007az,Lake:2006pp,Geralico:2012jt}. In this sense, the modified gravity sector provides an effective halo whose phenomenology is observationally indistinguishable, at the level of circular dynamics, from a standard dark matter distribution. Our construction also provides an example of how an effective fluid description can emerge from a fundamental action principle, being completely indistinguishable at the background level.

The deflection of light and gravitational lensing produced by Einstein cluster configurations were studied in Ref.~\cite{Boehmer:2007az}. There it was shown that, for geometries reproducing flat galactic rotation curves, the predicted bending angle differs only modestly from that of a standard pressureless dark matter halo: the Einstein cluster configuration generally produces a slightly smaller deflection angle for the same circular velocity profile. Although the difference is quantitatively small, it reflects the anisotropic stress inherent to the cluster description and therefore constitutes a potentially observable distinction between the two scenarios. In principle, sufficiently precise future lensing measurements could test this deviation and provide an independent observational probe of whether galactic halos are sourced by pressureless matter or by a halo of the Einstein cluster type.

We have performed a preliminary study of the radial stability of our system by considering time-dependent monopole perturbations of the vector field of the form $v_0 = \bar{v}_0 + \epsilon\, e^{i \omega t} v_{01}(r)$ and $v_1 = \bar{v}_1 + \epsilon\, e^{i \omega t} v_{11}(r)$, where $\epsilon$ is a small bookkeeping parameter, and the bar denotes the background solution. Similarly to other spherically symmetric systems with vector fields, such as the Reissner-Nordström metric, we find that monopole perturbations do not carry dynamical degrees of freedom and are purely constrained, effectively vanishing. This behavior indicates stability of the background against radial perturbations, as expected given that Einstein clusters are known to be stable under radial modes~\cite{Boehmer:2007az}. A full perturbative analysis of the coupled metric–vector system, including more general polar and axial modes, lies beyond the scope of this letter and represents an important direction for future work.

Relativistic realizations of MOND~\cite{1984ApJ...286....7B,Sanders:1996wk,PhysRevD.70.083509,Zlosnik:2006zu,Babichev:2011kq,Khoury:2014tka,Skordis:2020eui} typically share a common structural ingredient: the presence of a vector field constrained to be timelike through the introduction of a Lagrange multiplier. Such a vector field spontaneously breaks local Lorentz invariance by selecting a preferred local frame, which is central to implementing gravitational dynamics sensitive to acceleration scales. However, enforcing this condition through a Lagrange multiplier often leads to dynamical instabilities in the theory~\cite{Contaldi:2008iw}.
In the present model, an analogous structure emerges dynamically. Indeed, combining Eqs.~\eqref{eq:relationmetric}, \eqref{eq:v1}, \eqref{eq:gamma}, and \eqref{eq:final_relation_derivation}, one finds directly that the norm of the vector field is fixed to $V_\mu V^\mu = -1/2\pi$.
The vector field is therefore everywhere timelike, implying spontaneous breaking of local Lorentz symmetry through the selection of a preferred frame. Unlike in standard relativistic MOND constructions, however, no external constraint or Lagrange multiplier is required: the timelike character of the vector field follows entirely from the field equations themselves. The preferred frame structure is not imposed kinematically, but arises as an intrinsic property of the gravitational sector at galactic scales.
Because the vector field has fixed norm, its temporal component admits a direct physical interpretation~\cite{Zlosnik:2006zu}. In the weak-field regime, where the metric takes the form $g_{tt}\approx -(1+2\phi)$, one finds at leading order that $\sqrt{2\pi} v_0 \approx 1+\phi$, while $v_1$ vanishes at the same order. The temporal component of the vector field is therefore directly related to the Newtonian gravitational potential, in the sense that its departure from the constant background value tracks the weak-field gravitational potential experienced by non-relativistic matter.

Although the theory \eqref{eq:theory} is treated as an effective field theory, it would be important to investigate whether it admits well-defined time evolution, particularly in light of the findings in Refs.~\cite{Unluturk:2023qgk,Coates:2023dmz,Coates:2022qia,Silva:2021jya,Garcia-Saenz:2021uyv,Clough:2022ygm}. If it does not, it would be valuable to explore what kinds of higher-order operators or UV completions could restore well-defined time evolution.

Our analysis shows that the choice $\gamma = 1/4$ is uniquely singled out by the requirement that the theory exactly reproduces the dynamics of an Einstein cluster. At present, however, we have not identified an underlying symmetry principle or fundamental mechanism that selects this value a priori. It would therefore be of considerable interest to determine whether $\gamma = 1/4$ is distinguished by additional properties. Such an understanding could clarify whether this coupling is merely phenomenologically selected or reflects a more fundamental property of the theory.

\noindent {\bf \em Acknowledgments.} 
%
We thank Fethi M. Ramazano\u{g}lu for useful comments.
We thank Christos Charmousis, Astrid Eichhorn, Mokhtar Hassaine and Benjamin Knorr for useful discussions. P.F. would like to thank the theory department at CERN for their kind hospitality during the initial stages of this work.
Calculations were performed using the ``OGRe'' Mathematica package \cite{Shoshany:2021iuc}.
P.F. is funded by the Deutsche Forschungsgemeinschaft (DFG, German Research Foundation) under Germany’s Excellence Strategy EXC 2181/1 - 390900948 (the Heidelberg STRUCTURES Excellence Cluster).
The Center of Gravity is a Center of Excellence funded by the Danish National Research Foundation under grant No. 184.
We acknowledge support by VILLUM Foundation (grant no. VIL37766) and the DNRF Chair program (grant no. DNRF162) by the Danish National Research Foundation.
V.C.\ is a Villum Investigator.  
V.C. acknowledges financial support provided under the European Union’s H2020 ERC Advanced Grant “Black holes: gravitational engines of discovery” grant agreement no. Gravitas–101052587. 
Views and opinions expressed are however those of the author only and do not necessarily reflect those of the European Union or the European Research Council. Neither the European Union nor the granting authority can be held responsible for them.
This project has received funding from the European Union's Horizon 2020 research and innovation programme under the Marie Sklodowska-Curie grant agreement No 101007855 and No 101131233. 
This work is supported by Simons Foundation International \cite{sfi} and the Simons Foundation \cite{sf} through Simons Foundation grant SFI-MPS-BH-00012593-11.
\bibliography{References}

\end{document}